\documentclass[prl,twocolumn,showpacs,english,floatfix,preprintnumbers,tighten,epsfig,superscriptaddress]{revtex4}
\usepackage{amsmath}
\usepackage{amsfonts}
\usepackage{graphicx}
\usepackage{amssymb}
\usepackage{graphicx}
\usepackage{babel}

\makeatletter \providecommand{\boldsymbol}[1]{\mbox{\boldmath $#1$}}
\makeatother

\begin{document}

\title{Directional spatial structure of dissociated elongated molecular
condensates }
\author{Magnus \"{O}gren}
\affiliation{ARC Centre of Excellence for Quantum-Atom Optics, School of Physical
Sciences, University of Queensland, Brisbane, QLD 4072, Australia}
\author{C.~M. Savage}
\affiliation{ARC Centre of Excellence for Quantum-Atom Optics, Department of Physics,
Australian National University, Canberra ACT 0200, Australia}
\author{K.~V. Kheruntsyan}
\affiliation{ARC Centre of Excellence for Quantum-Atom Optics, School of Physical
Sciences, University of Queensland, Brisbane, QLD 4072, Australia}
\date{\today{}}

\begin{abstract}
Ultra-cold clouds of dimeric molecules can dissociate into quantum
mechanically correlated constituent atoms that are either bosons or
fermions. We theoretically model the dissociation of cigar shaped molecular
condensates, for which this difference manifests as complementary geometric
structures of the dissociated atoms. For atomic bosons beams form along the
long axis of the molecular condensate. For atomic fermions beams form along
the short axis. This directional beaming simplifies the measurement of
correlations between the atoms through relative number squeezing.
\end{abstract}

\maketitle

The difference between bosons and fermions is fundamental in physics.
Ultra-cold degenerate quantum gases enable new kinds of explorations and
applications of this difference. For example in Hanbury Brown and Twiss type
measurements bosonic $^{4}$He atoms are observed to be bunched while
fermionic $^{3}$He atoms are anti-bunched \cite{Jeltes}. Besides such local
spatial correlations, non-local correlations between pairs of atoms with
opposite momenta have also been observed for both fermionic atoms \cite%
{Greiner} and bosonic atoms \cite{Perrin-BEC-collisions}. Such atom pairs,
generated by dissociation of molecular-dimers \cite{Greiner}, or by
four-wave mixing \cite{Perrin-BEC-collisions}, are predicted to be entangled
\cite%
{Pu-Meystre-Duan-EPR,Yurovski,twinbeams,Vardi,EPR-diss,Zhao,SavageKheruntsyanSpatial,PerrinNJP}%
. However the existence of non-classical Einstein-Podolsky-Rosen type
correlations \cite{Pu-Meystre-Duan-EPR,EPR-diss} between atom pairs in these
systems remains to be experimentally confirmed.

We report an analysis of the effect of particle statistics on the
dissociation of highly anisotropic Bose-Einstein condensates (BECs) of
molecular dimers. When the dissociated atoms are bosons we confirm the
expected beaming of atoms along the long axis of the cigar shaped molecular
condensate by stimulated emission \cite{Vardi}. However, when the atoms are
fermions we find an unexpected beaming in the orthogonal direction, i.e.,
along the short condensate axis. The effect is due to Pauli blocking
followed by atom-atom recombination which reduces the atomic density along
the long axis. We also find enhanced relative number squeezing between the
atoms in the half-spaces occupied by the oppositely directed beams.

Directionality effects due to bosonic amplification have been observed
\cite{Ketterle,Kozuma} in superradiant light scattering from elongated
condensates \cite{MooreMeystre,Vardi,UysMeystre}. The fermionic counterpart
of the effect due to Pauli blocking has not been discussed before to the
best of our knowledge.

The effective quantum field theory Hamiltonian describing the system is
given, in a rotating frame, by \cite{JOptB1999-PRA2000}
\begin{equation}
\widehat{H}=\widehat{H}_{0}+\hbar \int {}d\mathbf{x}\left[
\sum\limits_{i=1,2}\Delta \widehat{\Psi }_{i}^{\dagger }\widehat{\Psi }%
_{i}-ig(\mathbf{x})\left( \widehat{\Psi }_{1}\widehat{\Psi }_{2}-\widehat{%
\Psi }_{2}^{\dagger }\widehat{\Psi }_{1}^{\dagger }\right) \right] .
\label{hamiltonian}
\end{equation}%
Here, $\widehat{\Psi }_{i}(\mathbf{x},t)$ ($i=1,2$) are the field operators
for the atoms, which are two different spin states of the same isotope of
mass $m$, and may be either bosonic or fermionic. The field operators
satisfy the respective commutation or anti-commutaion relations, $[\widehat{%
\Psi }_{i}(\mathbf{x},t),\widehat{\Psi }_{j}^{\dagger }(\mathbf{x}^{\prime
},t)]=\delta _{ij}\delta ^{2}(\mathbf{x}-\mathbf{x}^{\prime })$ and $\{%
\widehat{\Psi }_{i}(\mathbf{x},t),\widehat{\Psi }_{j}^{\dagger }(\mathbf{x}%
^{\prime },t)\}=\delta _{ij}\delta ^{2}(\mathbf{x}-\mathbf{x}^{\prime })$.
In the following we shall only consider two spatial dimensions, with $%
\mathbf{x}=(x,y)$. However we expect our results to be at least
qualitatively valid for three-dimensional systems.

The first term in Eq.~(\ref{hamiltonian}) describes the kinetic energy of
the atoms $\widehat{H}_{0}=\int d\mathbf{x}\sum\nolimits_{i=1,2}\hbar ^{2}|%
\mathbf{\nabla }\widehat{\Psi }_{i}|^{2}/2m$. The detuning $\Delta $ is
defined so that spontaneous dissociation of molecules corresponds to $\Delta
<0$, with $2\hbar |\Delta |$ being the total dissociation energy that is
converted into the kinetic energy of atom pairs \cite{PMFT,Jack-Pu}. For
molecules at rest, the dissociation primarily populates the resonant atomic
modes in the two spin states having equal but opposite momenta, $\hbar
\mathbf{k}_{1}=-\hbar \mathbf{k}_{2}$, with the absolute wavenumber equal to
$k_{0}=|\mathbf{k}_{1}|=|\mathbf{k}_{2}|=\sqrt{2m|\Delta |/\hbar }$.

In our analysis we use the undepleted molecular field approximation. It is
valid for short dissociation times during which the converted fraction of
molecules does not exceed about $10\%$ \cite{Savage,PMFT}. In the regime of
validity of the undepleted molecular approximation, the dissociation
typically produces low density atomic clouds for which the $s$-wave
scattering interactions are negligible \cite{Savage}; hence their absence from
our Hamiltonian. Additionally, the atom-molecule interactions can be
neglected if the respective interaction energy per atom is much smaller than
the total dissociation energy $2\hbar |\Delta|$. We have confirmed the
validity of our approximations (which improve with increasing $|\Delta |$)
in the case of dissociation into bosonic atoms by comparing the present
results with exact, first-principles simulations using the positive-$P$
representation \cite{Savage}. For the case of fermionic atoms,
first-principles simulations with multimode inhomogeneous molecular
condensates remain under development \cite{Corney-fermionic}, and such a
comparison is not presently possible.

We assume that the bosonic molecules are in a coherent state initially, with
the density profile $\rho _{\text{M}}(\mathbf{x})$ given by the ground state
solution of the Gross-Pitaevskii equation with an anisotropic harmonic trap.
We then have an effective, spatially dependent coupling
\begin{equation}
g(\mathbf{x})=\chi \sqrt{\rho _{\text{M}}(\mathbf{x})}.
\end{equation}%
The coupling coefficient $\chi $ \cite{PMFT} is responsible for coherent
conversion of molecules into atom pairs, e.g. via optical Raman transitions,
an rf pulse, or a Feshbach resonance sweep \cite%
{PDKKHH-1998-Superchemistry,Timmermans-JJ-1999,Holland,Feshbach-KKPD,Stoof-review-Julienne-review}%
. We assume that once the dissociation is switched on at time $t=0$%
, the trapping potential is switched off, so that the evolution is taking
place in free space. Thus the role of the trapping potential is reduced to
defining the initial shape of the molecular BEC.

The Heisenberg equations for the atomic fields in the Hamiltonian (\ref%
{hamiltonian}) are then:
\begin{align}
\frac{\partial \widehat{\Psi }_{1}(\mathbf{x},t)}{\partial t}& =i\left[
\frac{\hbar }{2m}\mathbf{\nabla }^{2}-\Delta \right] \widehat{\Psi }_{1}(%
\mathbf{x},t)\pm g(\mathbf{x})\widehat{\Psi }_{2}^{\dag }(\mathbf{x},t),
\label{Heisenberg-eqs} \\
\frac{\partial \widehat{\Psi }_{2}^{\dagger }(\mathbf{x},t)}{\partial t}& =-i%
\left[ \frac{\hbar }{2m}\mathbf{\nabla }^{2}-\Delta \right] \widehat{\Psi }%
_{2}^{\dagger }(\mathbf{x},t)+g(\mathbf{x})\widehat{\Psi }_{1}(\mathbf{x},t).
\notag
\end{align}%
The $+/-$ in the first equation correspond to bosonic/fermionic atoms.

Expanding in plane wave modes, $\widehat{\Psi }_{j}(\mathbf{x},t)=\int d^{2}%
\mathbf{k}\,\widehat{a}_{j}(\mathbf{k},t)\exp (-i\mathbf{k}\cdot \mathbf{x}%
)/2\pi $, where the operator amplitudes satisfy commutation or
anticommutation relations, $[\widehat{a}_{i}(\mathbf{k},t),\widehat{a}%
_{j}^{\dagger }(\mathbf{k}^{\prime },t)]=\delta _{ij}\delta ^{2}(\mathbf{k}-%
\mathbf{k}^{\prime })$ or $\{\widehat{a}_{i}(\mathbf{k},t),\widehat{a}%
_{j}^{\dagger }(\mathbf{k}^{\prime },t)\}=\delta _{ij}\delta ^{2}(\mathbf{k}-%
\mathbf{k}^{\prime })$, according to the underlying statistics. The partial
differential equations (\ref{Heisenberg-eqs}) reduce to a set of linear
ordinary differential equations,
\begin{gather}
\frac{d\widehat{a}_{1}(\mathbf{k},t)}{dt}=-i\Delta _{k}\widehat{a}_{1}(%
\mathbf{k},t)\pm \int \frac{d^{2}\mathbf{q}}{2\pi }\widetilde{g}(\mathbf{q}+%
\mathbf{k})\widehat{a}_{2}^{\dagger }(\mathbf{q},t),
\label{HeisenbergsEquation} \\
\frac{d\widehat{a}_{2}^{\dagger }(\mathbf{k},t)}{dt}=i\Delta _{k}\widehat{a}%
_{2}^{\dagger }(\mathbf{k},t)+\int \frac{d^{2}\mathbf{q}}{2\pi }\widetilde{g}%
(\mathbf{q}-\mathbf{k})\widehat{a}_{1}(-\mathbf{q},t),  \notag
\end{gather}%
where $\widetilde{g}(\mathbf{k})=\int d^{2}\mathbf{x}g(\mathbf{x})\,\exp (i%
\mathbf{k}\cdot \mathbf{x})/2\pi $ is the Fourier transform of $g(\mathbf{x})
$ and $\Delta _{k}\equiv \hbar k^{2}/\left( 2m\right) +\Delta $, where $k=|%
\mathbf{k}|$. For vacuum initial conditions the non-zero second-order
moments are the normal and anomalous densities; $n_{i}(\mathbf{k},\mathbf{k}%
^{\prime },t)\equiv \langle \widehat{a}_{i}^{\dagger }(\mathbf{k},t)\widehat{%
a}_{i}(\mathbf{k}^{\prime },t)\rangle $ and $m_{12}(\mathbf{k},\mathbf{k}%
^{\prime },t)\equiv \langle \widehat{a}_{1}(\mathbf{k},t)\widehat{a}_{2}(%
\mathbf{k}^{\prime },t)\rangle $. Higher-order moments can be obtained from
these second-order moments using Wick's theorem, as the Hamiltonian is
quadratic in the field operators in the undepleted molecular approximation.

In a finite quantization volume the wave-vector $\mathbf{k}$ is discrete and
the plane wave mode annihilation and creation operators may be organized
into a vector $\vec{\hat{a}}$. The Heisenberg equations (\ref%
{HeisenbergsEquation}) may then be written in vector-matrix form as $d\vec{%
\hat{a}}/dt = M \vec{\hat{a}}$, 
where $M$ is a matrix of complex numbers. Linearity ensures that the
solutions of these operator equations can be found by numerically computing
the matrix exponential $\exp(Mt)$. We used an $81 \times 81$ numerical
lattice in momentum space, which gave the same results as for a $61 \times
61 $ lattice.

\begin{figure}[tbp]
\includegraphics[height=2.25cm]{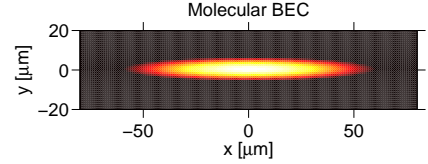}
\caption{(color online) Density in position space for the initial molecular
condensate. It corresponds to the ground state of a harmonic trap with
TF radii: $R_{\text{TF},x}\simeq 60$ $\protect\mu $m and $R_{\text{%
TF},y}\simeq 6$ $\protect\mu $m. }
\label{fig1}
\end{figure}
\begin{figure}[tbp]
\hspace{0.01cm}{\textbf{Bosonic atoms}} \hspace{1.8cm}{\textbf{Fermionic
atoms}} 
\includegraphics[height=3.25cm]{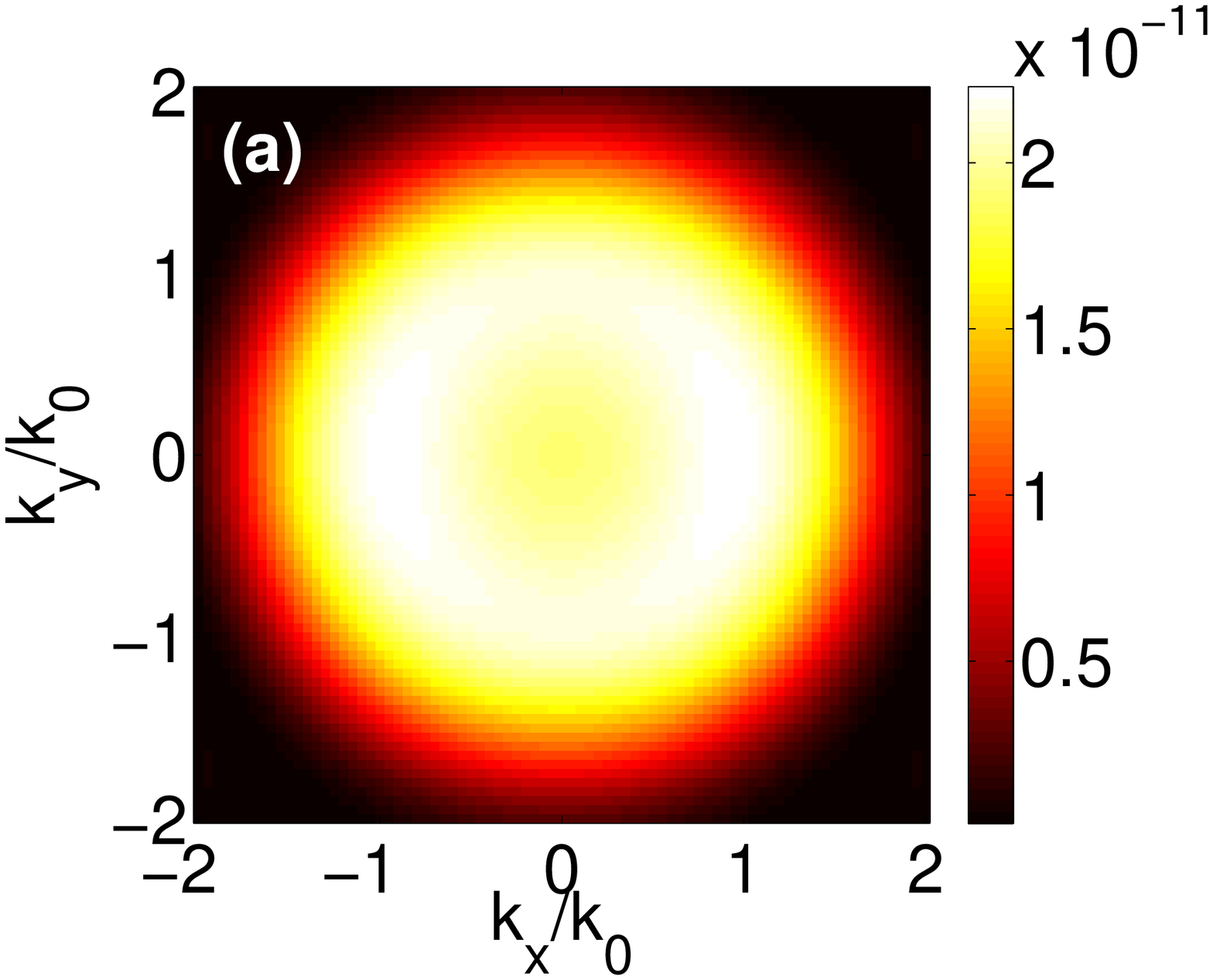}%
\includegraphics[height=3.25cm]{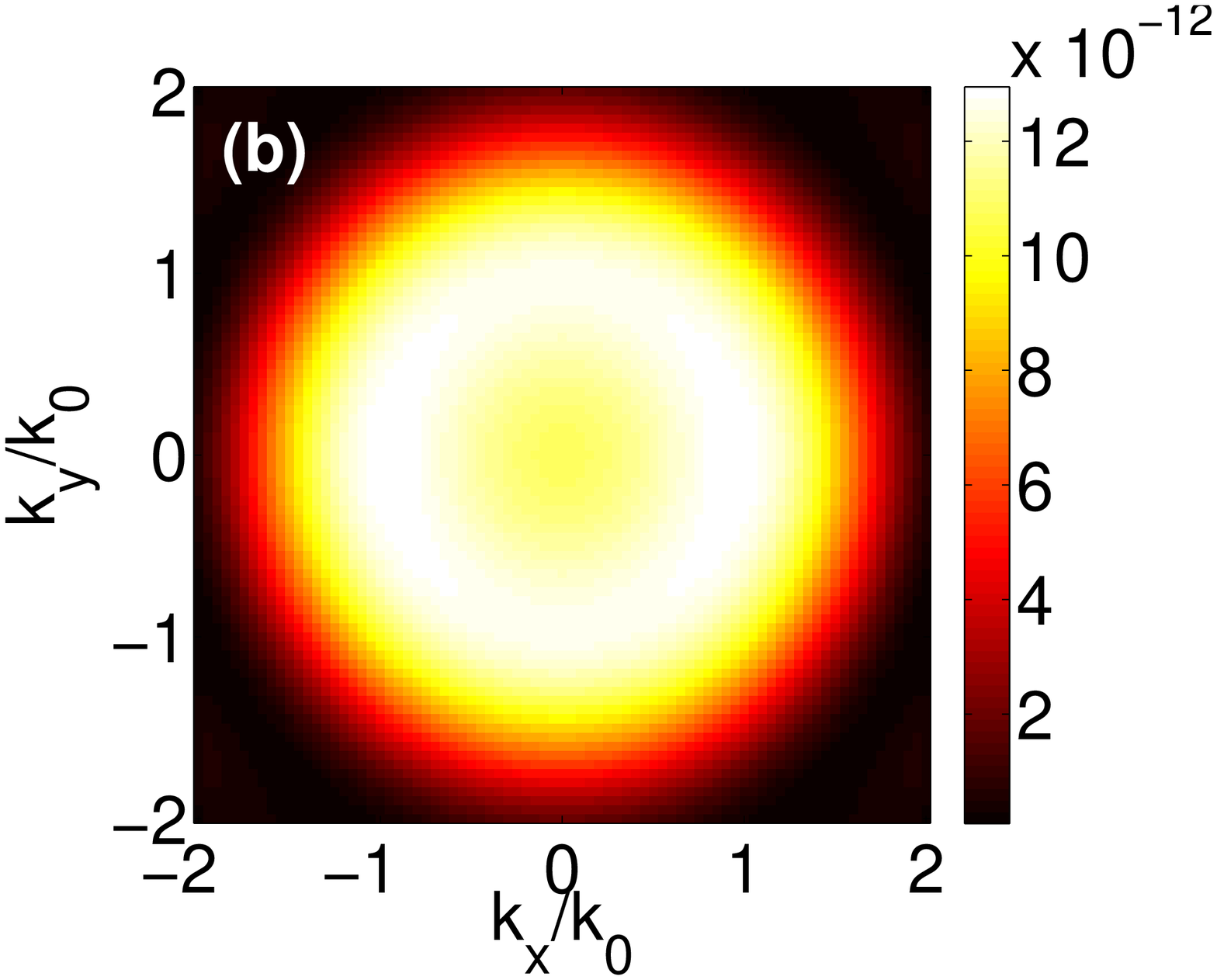} %
\includegraphics[height=3.25cm]{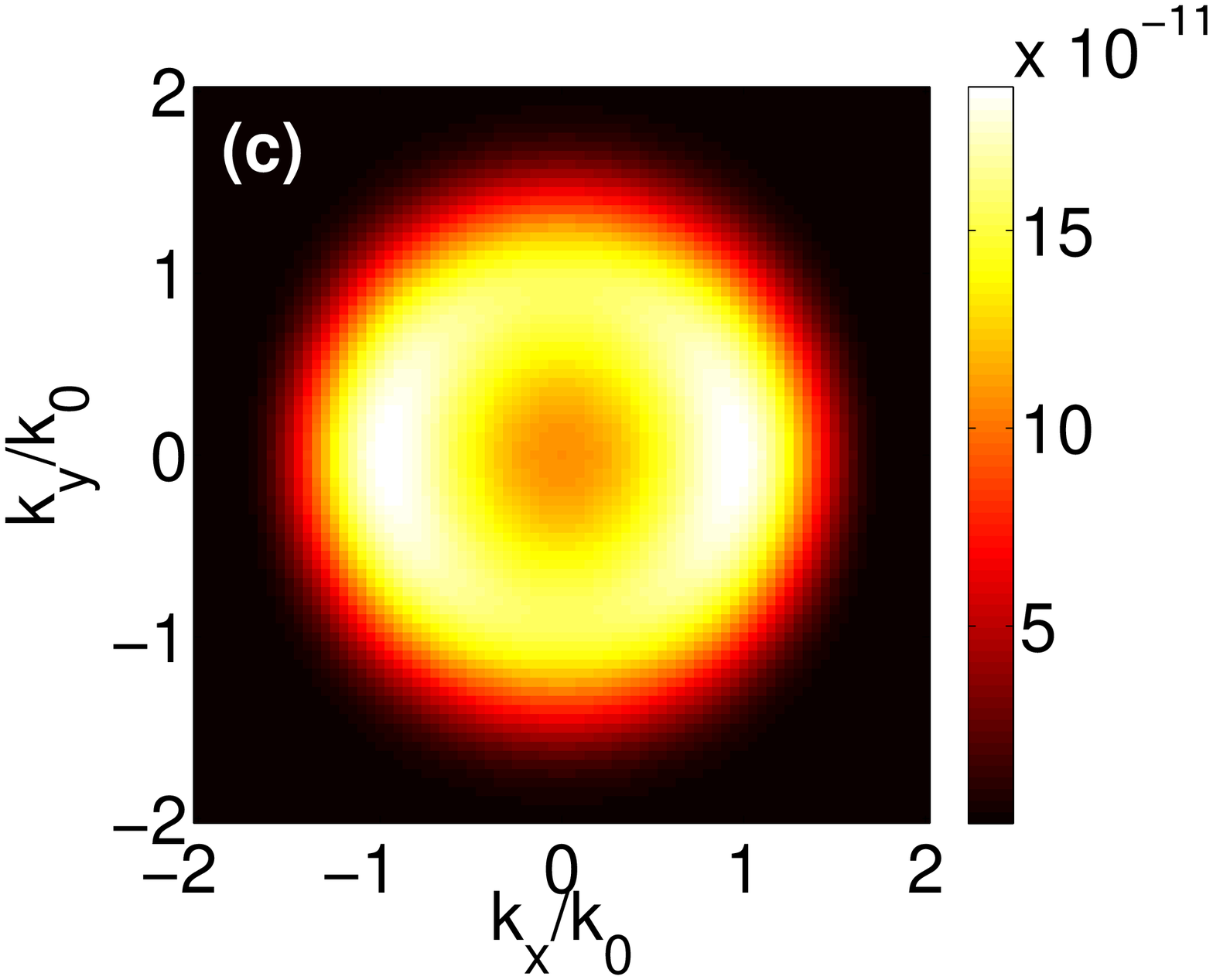}%
\includegraphics[height=3.25cm]{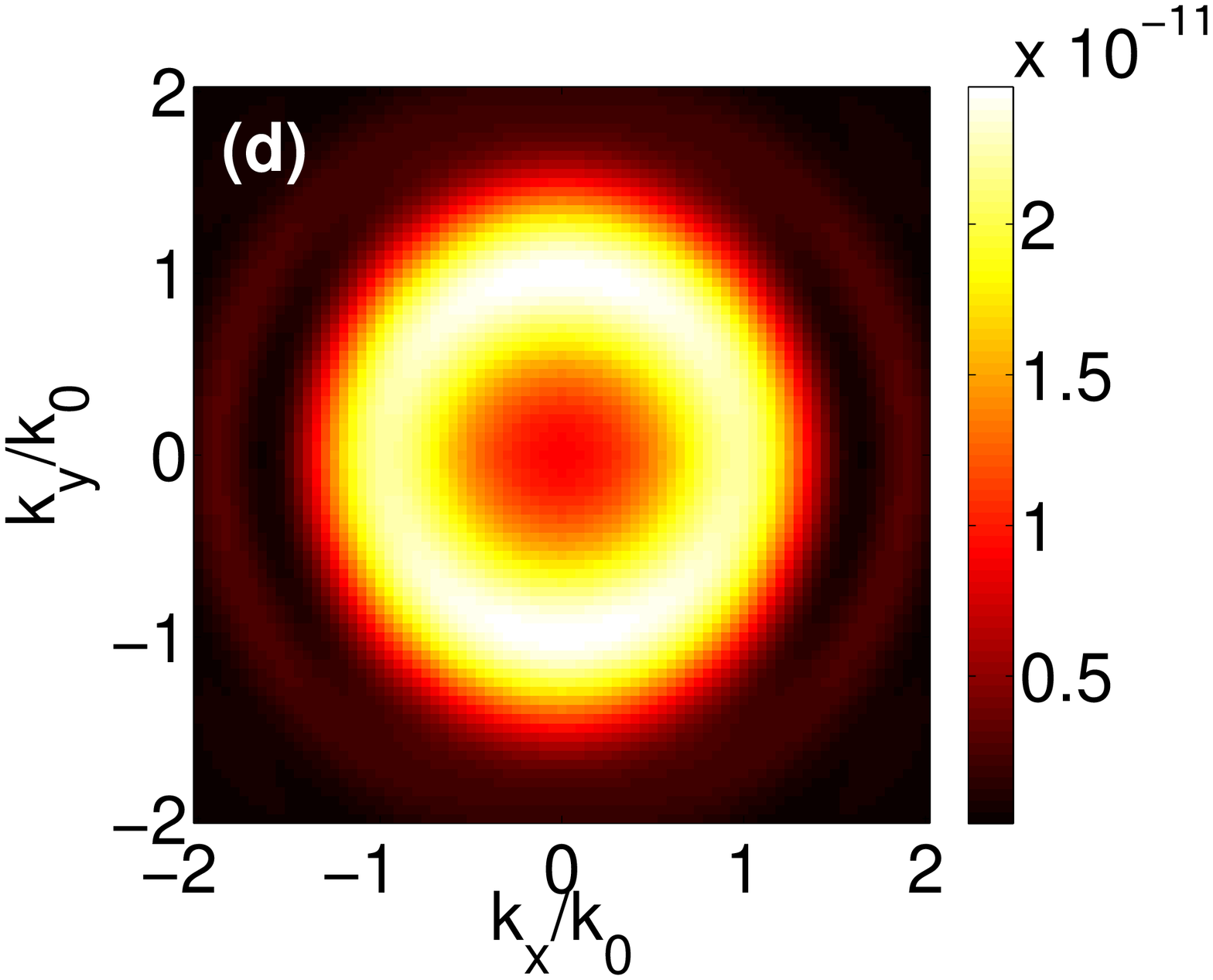} %
\includegraphics[height=3.25cm]{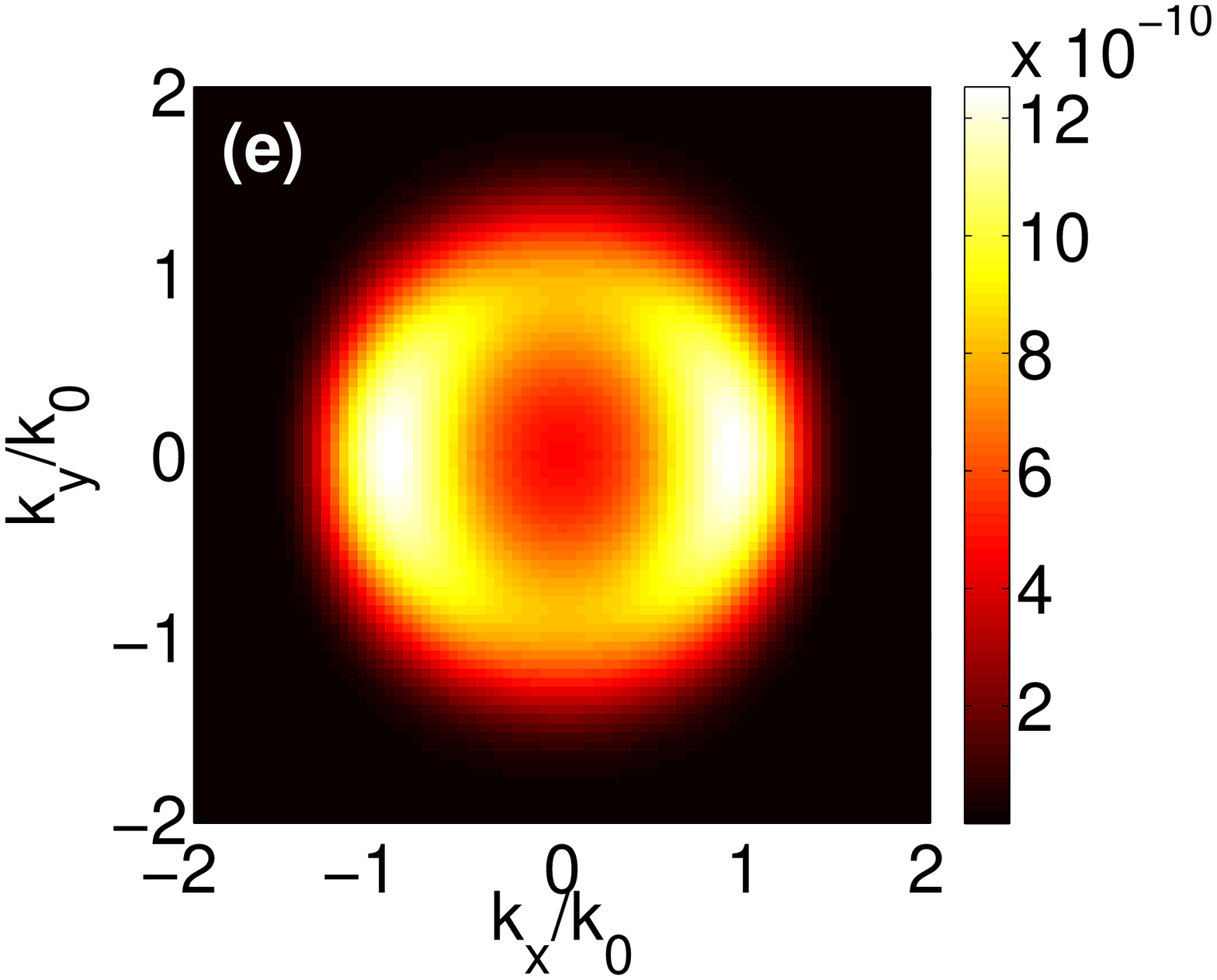}%
\includegraphics[height=3.25cm]{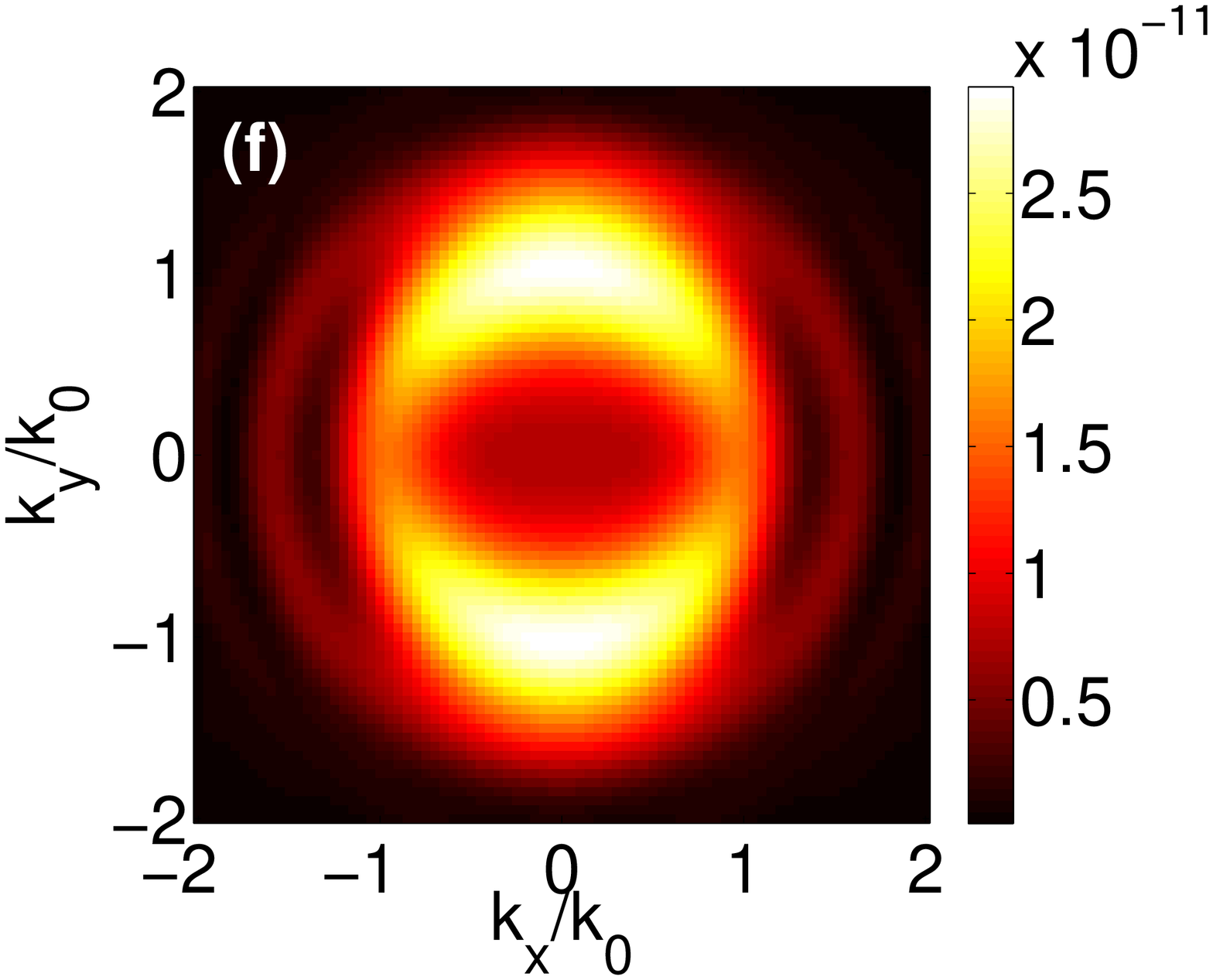}
\caption{(color online) Atomic density in momentum space (in units of m$^{2}$%
) at different times $t$ after the start of dissociation. The left
and right columns are for bosonic and fermionic atoms, respectively.
We have chosen the times $t_{1}/t_{0}=1.1$ for (a) and (b),
$t_{2}/t_{0}=2.2$ for (c) and (d), and $t_{3}/t_{0}=3.3$ for (e) and
(f). Here $t_{0}=1/(\protect\chi \protect\sqrt{\protect\rho
_{\text{M},0}})=2.5$ ms is the time scale.} \label{fig2}
\end{figure}

Here we explore the role of a strongly anisotropic system by assuming that
the initial molecular condensate was trapped in a harmonic potential with a
frequency along the $y$ axis $\omega _{y}$ that is ten times that along the $%
x$ axis; $\omega _{y}=10\omega _{x}$. We assume a molecule-molecule
scattering length such that the Thomas-Fermi (TF) approximation is valid.
The corresponding molecular density profile is $\rho _{\text{M}}=\rho _{%
\text{M},0}\left( 1-x^{2}/R_{\text{TF},x}^{2}-y^{2}/R_{\text{TF}%
,y}^{2}\right) $, where the $R_{\text{TF},i}$ are the TF-radii in
each direction. The parameter values we used are given in Ref.
\cite{Parameters}.

Fig.~\ref{fig1} shows the density of our initial two-dimensional molecular
condensate in position space. Fig.~\ref{fig2} shows the total density of the
dissociated atoms in momentum space, at three times as dissociation
progresses. The left column shows bosonic atoms, and the right column shows
fermionic atoms. Note that after sufficient time-of-flight expansion, these
momentum space distributions are reproduced in position space \cite%
{Greiner,SavageKheruntsyanSpatial}. At the earliest time there is little
difference between the bosons and fermions, and the width of the
distribution is determined by the energy-time uncertainty relation. Later,
the spatial distributions of the bosons and fermions develop quite
differently. For bosons, our results can be compared with
first-principles simulations using the positive-$P$ representation \cite%
{Savage}, in which the molecular field and its depletion is treated quantum
mechanically; the comparison is given in Ref. \cite{EPAPS} and shows good
agreement.

The highest densities of bosons develop along the long axis of the molecular
condensate. This is due to Bose stimulation, which in the undepleted
molecular field approximation leads to approximately exponential growth of
the atom number. In contrast, the highest densities of fermions occur in the
orthogonal direction; that is along the short axis of the initial molecular
condensate. Physically, the effect is due to Pauli blocking followed by the
reduction of the atomic density due to atom-atom recombination along the
long axis. The dynamics along the short axis (for sufficiently small $R_{%
\text{TF},y}$), on the other hand, does not reach the regime dominated by
saturation of mode populations and atom-atom recombination as the atoms
propagating along $y$ leave the molecular condensate earlier than those
propagating along $x$.

Mathematically, the difference is due to the sign difference on the
right-hand-side of the Heisenberg equations (\ref{Heisenberg-eqs}),
corresponding to bosons and fermions. As noted in Ref.~\cite{Fermidiss} the
sign difference determines $\sinh (\alpha _{k}\tau )$ population growth for
bosons, and $\sin (\alpha _{k}\tau )$ population oscillations for fermions;
where $\tau =t/t_{0}$ is a dimensionless time, with $t_{0}=1/(\chi \sqrt{%
\rho _{\text{M},0}})$, and $\alpha _{k}=\sqrt{1\pm t_{0}^{2}(\hbar
k^{2}/(2m)+\Delta )^{2}}$; here, the $+$ corresponding to fermions and the $%
- $ to bosons.

\begin{figure}[tbp]
\includegraphics[height=4.6cm]{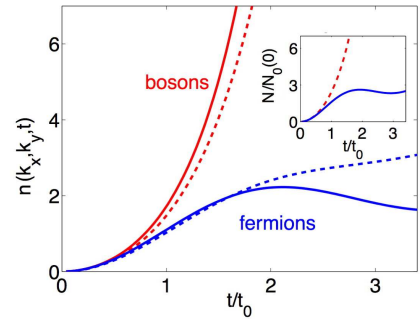}
\caption{(color online) Atomic density in momentum space at the position of
resonance ($\left\vert \mathbf{k}\right\vert =k_{0}$) along the two
different Cartesian directions, in units of $10^{-11}$ m$^{2}$. The two
upper (red) curves are for bosons which grow exponentially due to Bose
stimulation. The two lower (blue) curves are for fermions. Solid lines are
along the $k_{x}$-direction ($k_{x}=k_{0},k_{y}=0$), dashed lines are along
the $k_{y}$-direction ($k_{x}=0,k_{y}=k_{0}$). The inset shows the
fractional atom number (relative to the total initial number of molecules),
in units of tenths of a percent, for fermions (solid blue) and bosons
(dashed red).}
\label{fig3}
\end{figure}

The fermionic oscillations -- both in momentum space and in time -- can be
seen in Figs.~\ref{fig2} (d) and (f) and in Fig.~\ref{fig3}. As the
solutions of Ref.~\cite{Fermidiss} are for spatially homogenous systems they
do not quantitatively describe these minima for our inhomogeneous molecular
condensate. Nevertheless they predict aspects of the qualitative behavior,
such as the movement of the density minima and maxima to lower values of $%
k_{x}$ with increasing time; Figs.~\ref{fig2} (d) and (f). The single
density maximum along the $k_{y}$ axis is a result of the small TF
radius along that axis, $R_{\text{TF},y}=6$ $\mu $m. The atomic velocity at $%
|k/k_{0}|=1$ is $0.95$ mm s$^{-1}$, so during $t_{1}=1.1t_{0}=1.1(2.5$ ms)
the atoms travel $2.6$ $\mu $m, and during $t_{3}=3.3t_{0}$ they travel $7.8$ $%
\mu $m, which is greater than the TF radius $R_{\text{TF},y}$.
Hence a particular atom traveling in the $y$ direction typically interacts
with the molecular condensate for less than the time $t_{3}$. This accounts
for the similarity of Figs.~\ref{fig2} (d) and (f) along the $y$ axis.

We have used our solutions of the Heisenberg equations (\ref%
{HeisenbergsEquation}) to calculate the correlations between atom number
fluctuations in the two different spin states $1$ and $2$. We considered
momentum areas centered on the opposite resonant momenta, $\hbar \mathbf{k}%
_{0}$ and $-\hbar \mathbf{k}_{0}$, for the smallest and largest possible
areas in which the atoms are captured and their number is measured. The
smallest momentum areas are the numerical lattice areas, and the largest are
opposite halves of the momentum space. These are orientated differently for
fermions and bosons, guided by Figs.~\ref{fig2} (e) and (f), to capture one
beam in each half: for bosons the halves are split by the $k_{y}$ axis
and for fermions by the $k_{x}$ axis.

\begin{figure}[tbp]
\includegraphics[height=4.6cm]{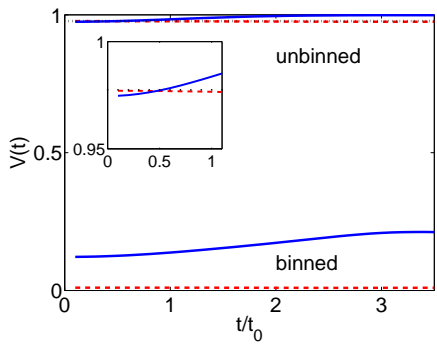}
\caption{(color online) Relative number variance according to
Eqs.~(\protect\ref{V-def1}) and (\protect\ref{V-def2}), raw
(unbinned) and binned, for fermions (solid blue) and bosons (dashed
red). The dotted
(black) line in the inset is the $t\ll t_{0}$ asymptote, Eq.~(\protect\ref%
{Analytic2DMDVk0-k0}).}
\label{fig4}
\end{figure}

For the case of the smallest momentum area (the area of the numerical
lattice cell, $\Delta k_{x}\Delta k_{y}$), we define the atom number
operators via $\hat{n}_{j,\pm \mathbf{k}_{0}}(t)=\hat{n}_{j}(\pm \mathbf{k}%
_{0},t)\Delta k_{x}\Delta k_{y}$, where $\mathbf{k}_{0}=k_{0}\mathbf{e}_{x}$
 for bosons and $\mathbf{k}_{0}=k_{0}\mathbf{e%
}_{y}$ for fermions, with $\mathbf{e}_{x}$ and $\mathbf{e}_{y}$ being the
Cartesian unit vectors. We quantify the correlations by the normalized
relative atom number variance,
\begin{equation}
V_{\mathbf{k}_{0},-\mathbf{k}_{0}}(t)=\langle \lbrack \Delta (\hat{n}_{1,%
\mathbf{k}_{0}}-\hat{n}_{2,-\mathbf{k}_{0}})]^{2}\rangle /\Delta _{\text{SN}%
},  \label{V-def1}
\end{equation}%
where $\Delta \hat{C}=\hat{C}-\langle \hat{C}\rangle $ is the fluctuation in
$\hat{C}$ and $\Delta _{\text{SN}}=\langle (\Delta \hat{n}_{1,\mathbf{k}%
_{0}})^{2}\rangle +\langle (\Delta \hat{n}_{2,-\mathbf{k}_{0}})^{2}\rangle $
defines the uncorrelated shot-noise level. Variance $V_{\mathbf{k}_{0},-%
\mathbf{k}_{0}}(t)<1$ implies squeezing of the relative number fluctuations
below the shot-noise level. The results (see Ref.~\cite{EPAPS} for details)
are shown in Fig.~\ref{fig4} (upper curves), where we see relatively small
degree of squeezing for this \textquotedblleft raw\textquotedblright\
(unbinned) case.

The squeezing in the relative number can be enhanced for larger counting
areas, which we refer to as binning \cite{SavageKheruntsyanSpatial}. For the
largest pair of bins coinciding with the left (L) and right (R) halves of
the momentum space for bosons, and with the bottom (B) and top (T) halves
for fermions, we can introduce particle number operators $\hat{N}_{i}^{\text{%
L}(\text{R})}$ and $\hat{N}_{i}^{\text{B}(\text{T})}$ ($i=1,2$). The
normalized variance of the relative number fluctuations between the spin
states $1$ and $2$ is defined similarly to Eq. (\ref{V-def1}),
\begin{equation}
V_{\text{LR(BT)}}(t)=\langle \lbrack \Delta (\hat{N}_{1}^{\text{R(T)}}-\hat{N%
}_{2}^{\text{L(B)}})]^{2}\rangle /\Delta _{\text{SN}}.  \label{V-def2}
\end{equation}
The results for $V_{\text{LR(BT)}}(t)$ (see Ref.~\cite{EPAPS} for details)
are shown in Fig.~\ref{fig4}; comparing the upper and lower curves, we see
that binning into momentum half-spaces reduces the relative number variance,
or increases the degree of squeezing, by a factor of about ten for fermions
and a hundred for bosons. This difference is due to the different momentum
uncertainties along the long and short spatial axes. The momentum
uncertainty is about ten times smaller along the long axis, which is the
bosonic beaming direction. Hence there are about ten times less correlated
pairs ending up in the same momentum half-space. For the fermions, on the
other hand, the relatively high momentum uncertainty in the beaming
direction produces more pairs in the same half-space, reducing the number
difference squeezing.

In the case of unbinned atom numbers the fermionic and bosonic cases have a
common short time asymptote that can be determined analytically using a
perturbative approach \cite{Ogren}. Applying it to our 2D system gives
\begin{equation}
V_{\mathbf{k}_{0},-\mathbf{k}_{0}}=1-2R_{TF,x}R_{TF,y}\Delta k_{x}\Delta
k_{y}/\left( 9\pi \right) \simeq 0.978,  \label{Analytic2DMDVk0-k0}
\end{equation}%
which agrees with our numerical results (see Fig. \ref{fig4}).

In summary, we have shown that dissociation of elongated condensates of
molecular dimers into atoms can produce qualitatively different geometrical
distributions for bosonic and fermionic atoms. These are in the form of
\textquotedblleft twin\textquotedblright\ beams in opposite half-spaces. The
squeezing of the relative atom number fluctuations between these beams can
be quite strong.

The authors acknowledge support from the Australian Research Council through
the ARC Centre of Excellence scheme. M\"{O} acknowledges IPRS/UQILAS.


\vspace{2cm}

\setcounter{MaxMatrixCols}{10}

\makeatletter \providecommand{\boldsymbol}[1]{\mbox{\boldmath $#1$}}
\makeatother

\textbf{Supplementary material for EPAPS}
\vspace{0.5cm}

1. \textit{Comparison with the positive-P simulations.} --- Here, we compare
the results of the present treatment using the undepleted molecular
approximation and exact first-principles simulations using the positive-$%
P $ representation \cite{SavageKheruntsyanSpatial, Savage}, in which the molecular depletion is taken
into account within the full quantum field theory Hamiltonian. In Fig.~\ref%
{fig1} we show the corresponding atomic densities for the case of bosonic
atoms. Figure~\ref{fig1}~(a) is the same one as Fig.~2~(e) of the main text,
while Fig.~\ref{fig1}~(b) is the corresponding positive-$P$ result. As we see
the agreement is very good. First principles simulations for fermionic atoms
using the Gaussian stochastic methods of Ref.~\cite{Corney-fermionic} are
still under development, but we expect a similarly good agreement as the
molecular depletion is a weaker effect in this case \cite{Fermidiss,PMFT}.

\begin{figure}[h]
\includegraphics[height=3.4cm]{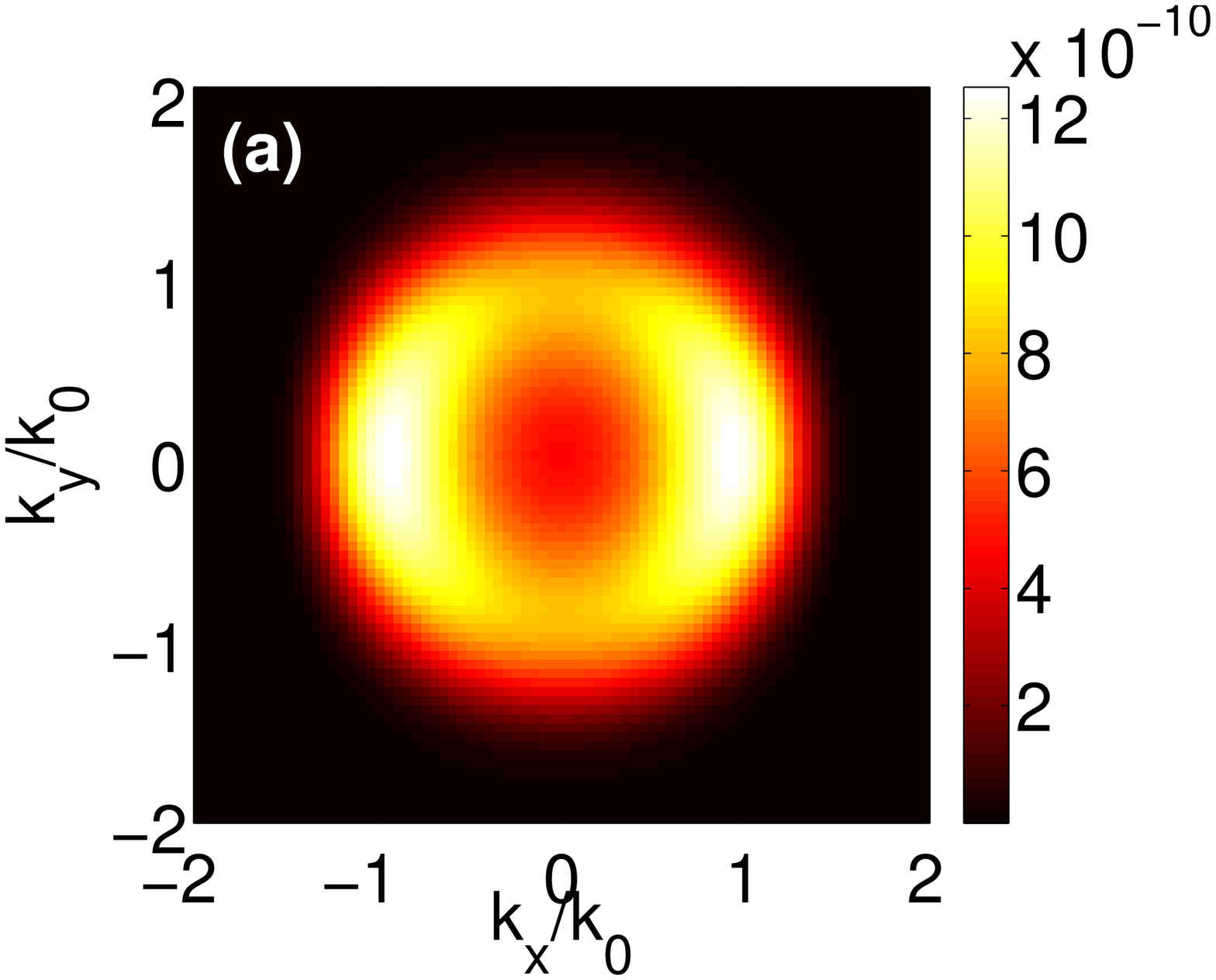}%
\includegraphics[height=3.4cm]{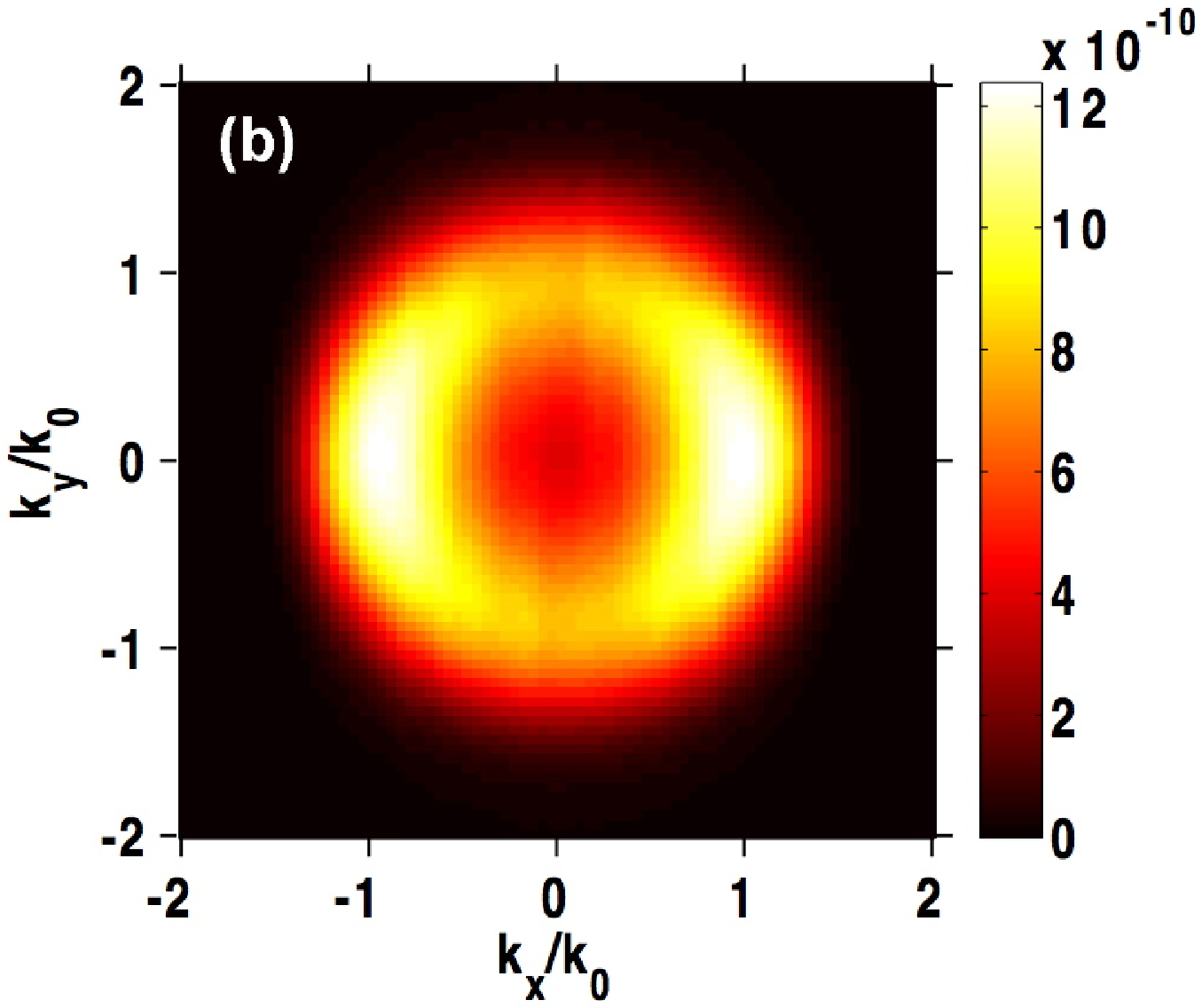}
\caption{Atomic density in momentum space (in units of m$^2$) at
$t_{3}/t_{0}=3.3$ within the undepleted molecular field
approximation (a) and using the positive-$P$ simulations with
molecular depletion (b).} \label{fig1}
\end{figure}

2. \textit{Shot noise and relative number variances.} --- For bosons the
uncorrelated shot-noise is determined by Poissonian statistics so that in
the case of the smallest counting area we have
\begin{eqnarray}
\Delta _{\text{SN}} &=&\langle (\Delta \hat{n}_{1,\mathbf{k}%
_{0}})^{2}\rangle +\langle (\Delta \hat{n}_{2,-\mathbf{k}_{0}})^{2}\rangle
\notag \\
&=&\langle \hat{n}_{1,\mathbf{k}_{0}}\rangle +\langle \hat{n}_{2,-\mathbf{k}%
_{0}}\rangle .
\end{eqnarray}%
For fermions, on the other hand, the shot-noise is always sub-Poissonian and
is given by \cite{Fermidiss}
\begin{eqnarray}
\Delta _{\text{SN}} &=&\langle (\Delta \hat{n}_{1,\mathbf{k}%
_{0}})^{2}\rangle +\langle (\Delta \hat{n}_{2,-\mathbf{k}_{0}})^{2}\rangle
\notag \\
&=&\langle \hat{n}_{1,\mathbf{k}_{0}}\rangle (1-\langle \hat{n}_{1,\mathbf{k}%
_{0}}\rangle )+\langle \hat{n}_{2,-\mathbf{k}_{0}}\rangle (1-\langle \hat{n}%
_{2,-\mathbf{k}_{0}}\rangle ).  \notag \\
&&\;
\end{eqnarray}

The relative number variance can be written in the following form \cite%
{SavageKheruntsyanSpatial, Savage,Ogren}:%
\begin{equation}
V_{\mathbf{k}_{0},-\mathbf{k}_{0}}(t)=\frac{\langle \hat{n}_{1,\mathbf{k}%
_{0}}\hat{n}_{1,\mathbf{k}_{0}}\rangle -\langle \hat{n}_{1,\mathbf{k}_{0}}%
\hat{n}_{2,-\mathbf{k}_{0}}\rangle }{\Delta _{\text{SN}}},
\end{equation}%
where we have taken into account that $\langle \hat{n}_{1,\mathbf{k}%
_{0}}\rangle =\langle \hat{n}_{2,-\mathbf{k}_{0}}\rangle \equiv n_{\mathbf{k}%
_{0}}$ and $\langle \hat{n}_{1,\mathbf{k}_{0}}\hat{n}_{1,\mathbf{k}%
_{0}}\rangle =\langle \hat{n}_{2,-\mathbf{k}_{0}}\hat{n}_{2,-\mathbf{k}%
_{0}}\rangle $. Note that $\langle \hat{n}_{1,\mathbf{k}_{0}}\hat{n}_{1,%
\mathbf{k}_{0}}\rangle =\langle \hat{n}_{1,\mathbf{k}_{0}}\rangle =n_{%
\mathbf{k}_{0}}$ for fermions. Applying Wick's theorem to calculate the
higher-order moments, and introducing the resonant anomalous density $m_{%
\mathbf{k}_{0}}\equiv \langle \hat{a}_{1,\mathbf{k}_{0}}\hat{a}_{2,-\mathbf{k%
}_{0}}\rangle $, where $\hat{a}_{j,\mathbf{k}}(t)\equiv \hat{a}_{j}(\mathbf{k%
},t)\sqrt{\Delta k_{x}\Delta k_{y}}$ we obtain%
\begin{equation}
V_{\mathbf{k}_{0},-\mathbf{k}_{0}}(t)=1-\frac{|m_{\mathbf{k}_{0}}|^{2}-n_{%
\mathbf{k}_{0}}^{2}}{n_{\mathbf{k}_{0}}}
\end{equation}%
for bosons, and
\begin{equation}
V_{\mathbf{k}_{0},-\mathbf{k}_{0}}(t)=1-\frac{|m_{\mathbf{k}_{0}}|^{2}}{n_{%
\mathbf{k}_{0}}(1-n_{\mathbf{k}_{0}})}.
\end{equation}%
for fermions. The variance $V_{\mathbf{k}_{0},-\mathbf{k}_{0}}(t)$ is then
calculated by numerically solving Eqs.~(4) of the main text.

For the case of the largest pair of counting areas, which we refer to as
binning, we consider the left (L) and the right (R) halves of the momentum
space for bosons, and the bottom (B) and top (T) halves for fermions.
Accordingly, we introduce particle number operators $\hat{N}_{j}^{\text{L}(%
\text{R})}$ and $\hat{N}_{j}^{\text{B}(\text{T})}$ ($j=1,2$), defined as $%
\hat{N}_{j}^{\text{L}}=\sum^{\text{L}}\hat{n}_{j,\mathbf{k}}$ (with a
similar convention for R, B, and T), where $\sum^{\text{L}}$and $\sum^{\text{%
R}}$ stand for double sums $\sum^{\text{L}}\equiv
\sum\nolimits_{k_{x}<0}\sum\nolimits_{k_{y}}$ and $\sum^{\text{R}}\equiv
\sum\nolimits_{k_{x}>0}\sum\nolimits_{k_{y}}$, whereas $\sum^{\text{B}%
}\equiv \sum\nolimits_{k_{x}}\sum\nolimits_{k_{y}<0}$ and $\sum^{\text{T}%
}\equiv \sum\nolimits_{k_{x}}\sum\nolimits_{k_{y}>0}$. The normalized
variance of the relative atom number fluctuations in the spin states $1$ and
$2$ is defined in Eq.~(6) of the main text, in which the uncorrelated
shot-noise level
\begin{equation}
\Delta _{\text{SN}}=\sum\nolimits^{\text{R}}\langle (\Delta \hat{n}_{1,%
\mathbf{k}})^{2}\rangle +\sum\nolimits^{\text{L}}\langle (\Delta \hat{n}_{2,%
\mathbf{k}})^{2}\rangle
\end{equation}%
for bosons is given by%
\begin{equation}
\Delta _{\text{SN}}=\sum\nolimits^{\text{R}}\langle \hat{n}_{1,\mathbf{k}%
}\rangle +\sum\nolimits^{\text{L}}\langle \hat{n}_{2,\mathbf{k}}\rangle
=\langle \hat{N}_{1}^{\text{R}}\rangle +\langle \hat{N}_{2}^{\text{L}%
}\rangle ,
\end{equation}%
while for fermions it is given by
\begin{equation}
\Delta _{\text{SN}}=\sum\nolimits^{\text{T}}\langle \hat{n}_{1,\mathbf{k}%
}\rangle (1-\langle \hat{n}_{1,\mathbf{k}}\rangle )+\sum\nolimits^{\text{B}%
}\langle \hat{n}_{2,\mathbf{k}}\rangle (1-\langle \hat{n}_{2,\mathbf{k}%
}\rangle ),
\end{equation}%
which we note is not same as $[\langle \hat{N}_{1}^{\text{T}}\rangle
(1-\langle \hat{N}_{1}^{\text{T}}\rangle )+\langle \hat{N}_{2}^{\text{B}%
}\rangle (1-\langle \hat{N}_{2}^{\text{B}}\rangle )]$. Applying Wick's
theorem to factorize the higher-order moments in Eq.~(6) of the main text,
we obtain
\begin{gather}
V_{\text{LR}}(t)=\frac{1}{\langle \hat{N}_{1}^{\text{R}}\rangle }\left[
\langle \hat{N}_{1}^{\text{R}}\rangle \right. \\
\left. +\sum\nolimits_{\mathbf{k}}^{\text{R}}\sum\nolimits_{\mathbf{k}%
^{\prime }}^{\text{R}}|\langle \hat{a}_{1,\mathbf{k}}^{\dagger }\hat{a}_{1,%
\mathbf{k}^{\prime }}\rangle |^{2}-\sum\nolimits_{\mathbf{k}}^{\text{R}%
}\sum\nolimits_{\mathbf{k}^{\prime }}^{\text{L}}|\langle \hat{a}_{1,\mathbf{k%
}}\hat{a}_{2,\mathbf{k}^{\prime }}\rangle |^{2}\right]  \notag
\end{gather}%
for bosons, and
\begin{gather}
V_{\text{BT}}(t)=\frac{1}{\langle \hat{N}_{1}^{\text{T}}\rangle
-\sum\nolimits_{\mathbf{k}}^{\text{T}}\langle \hat{n}_{1,\mathbf{k}}\rangle
^{2}}\left[ \langle \hat{N}_{1}^{\text{T}}\rangle \right. \\
\left. -\sum\nolimits_{\mathbf{k}}^{\text{T}}\sum\nolimits_{\mathbf{k}%
^{\prime }}^{\text{T}}|\langle \hat{a}_{1,\mathbf{k}}^{\dagger }\hat{a}_{1,%
\mathbf{k}^{\prime }}\rangle |^{2}-\sum\nolimits_{\mathbf{k}}^{\text{T}%
}\sum\nolimits_{\mathbf{k}^{\prime }}^{\text{B}}|\langle \hat{a}_{1,\mathbf{k%
}}\hat{a}_{2,\mathbf{k}^{\prime }}\rangle |^{2}\right]  \notag
\end{gather}%
for fermions. Here, we have used the fact that $\langle \hat{n}_{1,\mathbf{k}%
}\rangle =\langle \hat{n}_{1,-\mathbf{k}}\rangle =\langle \hat{n}_{2,\mathbf{%
k}}\rangle $, $\langle \hat{N}_{1}^{\text{R}}\rangle =\langle \hat{N}_{2}^{%
\text{L}}\rangle $, and $\langle \hat{N}_{1}^{\text{T}}\rangle =\langle \hat{%
N}_{2}^{\text{B}}\rangle $, due to symmetry considerations. The calculation
of the variances $V_{\text{LR}}(t)$ and $V_{\text{BT}}(t)$ can now be done
using the numerical solution of Eqs.~(4) of the main text.


\begin{thebibliography}{99}
\bibitem{Parameters} Molecular peak density $\rho _{%
\text{M},0}=3.2\times 10^{13}$ m$^{-2}$; number of molecules $N_{\text{M}%
}=2\times 10^{4}$; atomic mass $m=6.5\times 10^{-26}$ kg ($^{40}$K);
molecule-atom coupling strength $\chi =7.1\times 10^{-5}$ ms$^{-1}$;
and dissociation detuning $\Delta =-281$ s$^{-1}$, which is large
enough to ensure that the dissociation energy is larger than thermal
at nK temperatures. The momentum lattice had a spacing $\Delta
k_{x}=\Delta k_{y}=3.0\times 10^{4}$ m$^{-1}$, which is smaller than
the smallest width of the molecular momentum distribution.
\end{thebibliography}
\end{document}